\documentclass[12pt,twoside]{article}

\usepackage{cite}
\usepackage{amsmath,amsfonts,amssymb,amsthm,mathabx}
\usepackage{times}
\usepackage{url}
\usepackage{hyperref}
\usepackage{tensor}
\usepackage{color}
\usepackage{caption}
\usepackage{enumitem}
\usepackage{graphicx}
\usepackage{array}
\usepackage{amsbsy}
\usepackage{appendix}
\usepackage{mathrsfs}
\usepackage{cancel}
\usepackage{yfonts}
\usepackage{inputenc}
\usepackage{chngcntr}
\usepackage{xcolor}
\usepackage{soul}
\usepackage{cancel}
\usepackage{pgfplots}
\usepackage[font=footnotesize,labelfont=bf]{caption}
\usepackage{subcaption}

\pgfplotsset{compat=1.16}
\usepackage{float}

\advance\textheight\topskip
\numberwithin{equation}{section} \makeatletter
\@addtoreset{equation}{section}

\begin{document}
	
	\title{Phase transition of photons and gravitons in a Casimir
		box}
	
	\author{Ankit Aggarwal, Glenn Barnich}
	
	%\date{}
	
	\def\mytitle{Phase transition of photons and gravitons in a Casimir
		box}
	
	\pagestyle{myheadings} \markboth{\textsc{\small A.~Aggarwal, G.~Barnich}}
	{\textsc{\small Phase transition in a Casimir box}}
	
	\addtolength{\headsep}{4pt}
	
	\begin{centering}
		
		\vspace{1cm}
		
		\textbf{\Large{\mytitle}}
		
		\vspace{1.5cm}
		
		{\large Ankit Aggarwal$^{a,b,c}$, Glenn Barnich$^a$}
		
		\vspace{1cm}
		
		\begin{minipage}{.9\textwidth}\small \it \begin{center}
				$^a$Physique Th\'eorique et Math\'ematique \\ Universit\'e libre de Bruxelles
				and International Solvay Institutes\\ Campus Plaine C.P. 231, B-1050
				Bruxelles, Belgium
			\end{center}
		\end{minipage}
		
		\vspace{.5cm}
		
			\begin{minipage}{.9\textwidth}\small \it \begin{center}
				$^b$	Institute for Theoretical Physics, TU Wien, Wiedner Hauptstraße~8-10, A-1040 Vienna, Austria
			\end{center}
		\end{minipage}
	
		\vspace{.5cm}
		
		\begin{minipage}{.9\textwidth}\small \it \begin{center}
				$^c$Institute for Theoretical Physics Amsterdam and Delta
				Institute for Theoretical Physics, University of Amsterdam,
				Science Park 904, 1098 XH Amsterdam, The Netherlands 
			\end{center}
		
		\end{minipage}
		
		\vspace{.5cm}
		
		E-mail:
		\href{mailto:aggarwal@hep.itp.tuwien.ac.at}{aggarwal@hep.itp.tuwien.ac.at},
		\href{mailto:gbarnich@ulb.ac.be}{gbarnich@ulb.ac.be}
		
	\end{centering}
	
	\vspace{1cm}

	\begin{center}
		\begin{minipage}{.9\textwidth} \textsc{Abstract}. A first order phase
			transition for photons and gravitons in a Casimir box is studied
			analytically from first principles with a detailed understanding of symmetry
			breaking due to the boundary conditions. It is closely related to
			Bose-Einstein condensation and accompanied by a quantum phase transition
			whose control parameter is the chemical potential for optical helicity.
		\end{minipage}
	\end{center}

	\vfill
	\thispagestyle{empty}
	
	\newpage
	
	\section{Introduction}
	\label{sec:introduction}
	When deriving the partition function for a free gas of photons, the black body
	result, the precise boundary conditions one imposes on the electric field and
	the vector potential are irrelevant in a high temperature/large volume limit.
	This is no longer so at low temperature/small volume since the details of the
	spectrum depend crucially on boundary conditions. In his review on quantum
	fields in curved space, DeWitt~\cite{DeWitt:1975ys} starts by discussing
	perfectly conducting boundary conditions for the electromagnetic field because
	they are consistent with gauge invariance and physically well-motivated. In
	the case of (linearized) gravity, the first part of the problem, finding
	boundary conditions that are consistent with gauge invariance, has been
	stressed for instance in~\cite{Avramidi1999,Witten:2018lgb} (see also
	references therein). The question whether a suitable box capable of confining
	and thermalizing gravitons does exist in principle has an interesting and
	controversial history%
	~\cite{Smolin1984,Smolin1985,garfinkle1985possibility,Dell1987,Padmanabhan_2003}.
	A lesson that one may infer from these discussions is that, even though it
	might not be possible to construct such a box out of standard matter, one
	should more generally think about spacetime boxes as surfaces where to impose
	boundary conditions: shells in black hole geometries as in Garfinkle and Wald,
	or black hole horizons where the boundary conditions on the electromagnetic
	field are duality symmetric in the analysis of Damour and
	Znajek~\cite{1978MNRAS.185..833Z,Damour:1978cg}, half way between perfectly
	conducting electric or magnetic ones (MIT bag boundary conditions). Addressing
	the problem of electromagnetic or perturbative gravitational fields in a
	realistic or physically relevant spacetime box that would consist for instance
	of a black horizon on the one hand and future null infinity on the other,
	requires one to deal simultaneously with quantum fields in curved spacetime
	and with non-trivial boundary conditions.
	
	The point of view we have taken in a series of papers on the subject is the
	following: the simplest, physically motivated, boundary conditions on the
	electromagnetic field that are compatible with gauge invariance are perfectly
	conducting electric boundary conditions in a slab geometry, i.e., those of the
	(idealized, finite-temperature) Casimir effect. The theoretical ``observable"
	that one might want to study in such a ``Casimir box'' is the partition
	function. Because of the simplicity of the background (Minkowski space-time)
	and the geometry (parallel plates along a single spatial direction), the
	partition function may be computed analytically as a function of the only
	relevant dimensionless parameter $b=\frac{\beta}{2a}$ where $a$ is the
	separation of the plates. This partition function has an intriguing inversion
	symmetry~\cite{Brown:1969na} $b\leftrightarrow 1/b$ that allows one to relate
	the high temperature/large volume expansion to the small temperature/small
	volume behavior. Our first analysis of the problem was devoted to the question
	of modular invariance in higher dimensions (see
	e.g.~\cite{Cappelli:1988vw,Shaghoulian:2015kta}), i.e., the question of what
	additional observable (coupled with an purely imaginary chemical potential) is
	needed in the electromagnetic case to generalize temperature inversion
	symmetry to full modular invariance as in the case of the free massless boson
	in $1+1$ dimensions~\cite{Alessio:2020okv}. In our second paper on the
	subject, we have provided explicit boundary conditions for linearized gravity
	in such a Casimir box that are consistent with gauge invariance and are
	in-line with the analysis in~\cite{Avramidi1999,Witten:2018lgb} which were
	used to show that the partition function is identical to that of photons in
	the same box~\cite{Alessio2021}. As has been emphasized, this theoretical
	exercise has nothing to add to the problem of whether such a box may in
	principle be constructed for gravitons or not. What it shows very clearly
	though is that in such a box, the entropy scales like the volume at high
	temperature and like the area of the plates at low temperature. One may thus
	wonder how, in this context of free gauge field theories subjected to
	non-trivial boundary conditions, one may trace this transition from the high
	to the low temperature behavior. In this note, this will be done based on our
	previous computations that will not be repeated here, through the coupling of
	an additional observable with real rather than purely imaginary chemical
	potential. Indeed, it has been argued convincingly in the last of the
	references~\cite{ROBERGE1986734,Alford:1998sd,PhysRevD.75.025003} devoted to
	the passage from imaginary to real chemical potentials that the former yield
	partition functions with interesting analycity properties while the latter
	gives access to the structure of phase transitions. In the case of photons,
	the analysis that we present below could have been done a long time ago
	starting with the results of~\cite{Brown:1969na}. 
	{This theoretical exercise is universal in the sense that it relies merely
		on free field theory, boundary conditions and standard equilibrium statistical
		physics. It is not directly concerned with concrete experimental setups,
		non-linearities, or non-equilibrium physics in where
		Bose-Einstein condensate of photons has been already discussed ~(see e.g.~\cite{Carusotto2013} for an
		extensive review).}
	
	In other words, a natural question when studying Bose-Einstein condensation of
	an ideal bose gas is whether such a phase transition also occurs in the case
	of photons or gravitons {in an idealized setup}. In the relativistic case, both particle and
	anti-particles have to be taken into account. The free model that has been
	solved analytically is a complex massive scalar field with chemical potential
	for $U(1)$ charge turned on~\cite{Haber1981,Haber:1981ts,Singh1984}. In this case, the critical
	behavior is different in the high and low temperature regimes. Furthermore,
	the critical temperature vanishes in two spatial dimensions. The usual
	argument why there is no such effect for photons is that they disappear into
	the walls and that there is no good conserved quantum number. Nevertheless,
	photon Bose-Einstein condensation~\cite{Klaers2010} has been observed recently in a
	cavity with curved mirrors where the system behaves effectively as a massive
	gas in two dimensions with a confining potential. In this context,
	polarization effects have been studied in~\cite{PhysRevA.96.043844}.
	
	The purpose of this paper is to provide a detailed theoretical understanding
	of the main mechanism of this phase transition, independently of the details
	of the experimental setup. The appropriate context turns out to be that of
	the Casimir effect~\cite{casimir1948attraction} at finite temperature~\cite{fierz_attraction_1960,Mehra:1967wf}, (see also
	e.g.~\cite{Brown:1969na,Dowker_1976,BALIAN1978165,10.1143/PTP.75.262,%
		Ambjorn:1981xw,Ambjorn:1981xv,Plunien:1987fr,Lutken:1988ge,Ford:1988gt,%
		PhysRevD.40.4191}
	and~\cite{Plunien:1986ca,Nesterenko:2005xv,Bordag:2009zzd}
	for reviews), that is to say, a photon or graviton gas confined between two
	perfectly conducting parallel plates. We will show that a first order phase
	transition, closely related to Bose-Einstein condensation, can be studied
	analytically from first principles by identifying the appropriate quantum
	number as optical helicity. This includes an understanding of how the symmetry
	between the two helicity states of the photon or gravitons is broken through
	the boundary conditions. A related breaking of chiral symmetry for photons
	through background curvature rather than through boundary conditions has been
	recently proposed in~\cite{Agullo:2016lkj,Rio2020}.
	
	Contrary to the massive scalar field case, we are dealing here with a finite
	size effect that depends on the precise boundary conditions. At low
	temperature, this is accompanied by a quantum phase transition of the same
	type as that studied in finite size systems \cite{fisher1978cr}, see e.g.~\cite{sachdev_2011,Gambassi:2009zz,Dantchev2022}
	for reviews. Our strategy will be to first do a detailed analysis of the
	problem for photons before generalising to the case of gravitons using the
	results of \cite{Alessio2021}. In Section \ref{sec:optical-helicity}, we introduce optical helicity as an
	observable central to the phase transition discussion. In Section \ref{sec:spectrum}, we
	study the spectrum of the electromagnetic field in a Casimir box and show that
	it is equivalent to that of a scalar field. The exact partition function is
	given in Section \ref{sec:partition-function} in low and high temperature regime. Section \ref{sec:bose-einst-cond} studies
	the Bose-Einstein condensation of a scalar field in dimensions greater than
	three. Section \ref{sec:behav-at-crit} shows that for the relevant case of three dimensions there
	is no Bose Einstein condensation for photons but there is a quantum phase
	transition. In Section \ref{sec:gravitons}, we argue that all the results for photons carry
	over to the case of gravitons with appropriate boundary conditions. We briefly
	comment on the question of physical realizability of our graviton boundary
	conditions. We end with some future directions in Section \ref{sec:discussion}.
	
	\section{Optical helicity}
	\label{sec:optical-helicity}
	
	In order to discuss the critical behavior of photons and gravitons in
	a Casimir box, an observable like occupation number for the ideal Bose
	gas or $U(1)$ charge for the relativistic complex scalar field is
	needed.
	
	In empty space the electromagnetic Hamiltonian is the superposition of
	harmonic oscillator Hamiltonians
	\begin{equation}
		\label{eq:76}
		\hat H=\sum_{\vec k}\hat{\mathcal{H}}_{\vec k},\quad \hat{\mathcal{H}}_{\vec k}=
		k(\hat a^{\alpha\dagger}_{\vec k}\delta_{\alpha\beta}\hat a^\beta_{\vec k}+\frac 12), 
	\end{equation}
	where $k=\sqrt{k_ik^i}$, $ i\in {1,2,3} $ and the index $\alpha=1,2$ denotes the two polarizations,
	which we take to be circular. Each of the individual Hamiltonians is invariant
	under transformations generated by the hermitian operators
	\begin{equation}
		\label{eq:77}
		\hat{\mathcal U}^A_{\vec k}=\hat a_{\vec
			k}^{\alpha\dagger}\sigma^A_{\alpha\beta}
		\hat a^\beta_{\vec k},\quad
		[\hat{\mathcal U}^A_{\vec k},\hat{\mathcal{H}}_{\vec k}]=0, 
	\end{equation}
	where $\sigma^0_{\alpha\beta}$ is the unit matrix and $\sigma^i_{\alpha\beta}$ are
	the Pauli matrices. We are interested here in $\hat {\mathcal{U}}^3_{\vec k}$, which counts
	the difference of the number of helicity $+1$ and helicity $-1$ photons,
	\begin{equation}
		\label{eq:79}
		\hat{\mathcal{U}}^3_{\vec k}=\hat a_{\vec k}^{1\dagger}\hat a^{1}_{\vec k}-
		\hat a_{\vec k}^{2\dagger}\hat a^{2}_{\vec k}. 
	\end{equation}
	More precisely, the relevant observable is
	\begin{equation}
		\hat	D=\sum_{\vec k} \hat{\mathcal U}^3_{\vec
			k}.\label{eq:25}
	\end{equation}
	
	In empty space where the potentials $\vec A,\vec Z$ for magnetic and
	electric fields $\vec B=\vec\nabla\times \vec A$,
	$\vec E=\vec\nabla\times\vec Z$ may be assumed to be transverse, the
	spacetime expression for $D$ is
	$\frac 12 \int d^3x (\vec A \cdot \vec B+\vec Z\cdot \vec E)$. It is
	related to the ``zilch'' \cite{Lipkin1964,Morgan1964,Kibble1965},
	generates duality rotations \cite{doi:10.1119/1.1971089,Deser1976}
	and is called optical helicity. Note however that this is not the
	spacetime expression of $D$ for the perfectly conducting boundary
	conditions that we consider below (see also \cite{Chernodub_2018} for
	related considerations in the cylindrical case).

	\section{The spectrum}
	\label{sec:spectrum}
	
	A Casimir box consists of the space between two parallel conducting plates,
	taken here normal to the $x^3$ axis, and separated by a distance $a$; we denote this space by $V$. When
	expressed in terms of the vector potential $\vec A$ and its conjugate momentum
	$\vec \pi=-\vec E$, perfectly conducting boundary conditions, 
	\begin{equation}
		(\vec E\times
		\vec n)|_{\partial V}=0=(\vec B\cdot\vec n)|_{\partial V}\label{eq:12}
	\end{equation}
	with $\vec n$ the normal to the boundary, are satisfied in Coulomb gauge if
	$A^a,E^a$ $a=1,2$ and $A^3,E^3$ obey Dirichlet and Neumann conditions
	respectively. This implies that $k_3=\frac{\pi n_3}{a}$, $n_3\in\mathbb N$ while
	$k_a=\frac{L_a n_a}{2\pi}$ where we take $L_a$ large and sums over $n_a\in
	\mathbb Z$ become $L_1L_2/(2\pi)^2$ times integrals over $k_a$.
	
	If $k_\perp=(k_ak^a)^{\frac 12}$ and $\epsilon^{ab}$ are completely
	anti-symmetric with $\epsilon^{12}=1$, linear polarization vectors
	adapted to the geometry of the problem are \cite{Alessio2021,Alessio:2020okv}
	\begin{equation}
		\begin{split}
			& e_H^a=k_\perp^{-1}\epsilon^{ab}k_b,\quad
			e_H^3=0,\\ & e_E^a=(k_\perp k)^{-1}k^ak_3,\ e_E^3=-(k_\perp k)^{-1}k_3^2,
		\end{split}
		\label{eq:14}
	\end{equation}
	with oscillators $a^H_{k},a^E_k$ for $H$ and $E$ modes respectively. When
	$k_3=0$, there are only $E$ but no $H$ modes, $a^H_{(k_a,0)}=0$. \footnote{$a_{(k_a,0)}$ refers to oscillators with $k_3=0$ and arbitrary transverse momenta $k_a$.} Circular
	polarization vectors are $e_\sigma^i=\frac{1}{\sqrt 2}(e_H^i+\sigma e_E^i)$,
	with $\sigma=\pm 1$ so that $a^\sigma_{k}=\frac{1}{\sqrt 2}(a^H_k-i\sigma a^E_k)$
	and the Hamiltonian is
	\begin{equation}
		\label{eq:2}
		H=\sum_{n_a}k_\perp a^{E*}_{(k_a,0)}a^{E}_{(k_a,0)}+\sum_{n_a,n_3>0,\sigma}k a^{\sigma*}_ka^\sigma_k,
	\end{equation}
	with the understanding that particle zero modes are dropped, while optical
	helicity becomes
	\begin{equation}
		\label{eq:3}
		D=\sum_{n_a,n_3>0,\sigma} \sigma a^{\sigma*}_ka^\sigma_k.
	\end{equation}
	
	In order to streamline the computation of the partition function, it is useful,
	but not essential, to consider a reformulation in terms of a massless scalar
	field $\phi(x)$ with periodic boundary conditions on the double interval $x^3\in
	[-a,a]$ of length $L_3=2a$ \cite{Alessio:2020okv,Alessio2021}. The
	oscillators associated to $\phi$ are related to those of the $E$ and $H$ modes
	through  (see appendix A.6 of \cite{Alessio2021})
	\begin{equation}
		\label{eq:4}
		a_k=\frac{1}{\sqrt 2}(a^E_k-ia^H_k),\ n_3\neq 0,\quad a_{(k_a,0)}=a^E_{(k_a,0)}.
	\end{equation} 
	This map is a canonical transformation in the sense that it preserves the
	Poisson brackets of the oscillators and maps the Hamiltonian in \eqref{eq:2} 
	to the one of a massless scalar,
	\begin{equation}
		\label{eq:5}
		H=\frac 12 \sum_{n_i}ka^*_ka_k.
	\end{equation}
	In these terms, the observable is
	\begin{equation}
		\label{eq:6}
		D=\sum_{n_i}{\rm sgn}(k_3)a^*_ka_k, 
	\end{equation}
	where ${\rm sgn}(k_3)$ is the sign of $k_3$.
	
	For the purpose of computing the partition function, one may then
	consider with no additional effort the more general case of a massless
	scalar field on a spacetime manifold with $d-1>0$ large spatial
	dimensions and one small spatial dimension,
	$\mathbb R^{d-1}\times \mathbb S^1_{L}\times \mathbb S^1_{\beta}$,
	with $D=\sum_{n_i}{\rm sgn} (k_d)a^*_ka_k$. The result for a photon
	gas in a Casimir box is obtained by setting $d=3$, and $L=2a$. This
	allows us to show that standard Bose-Einstein condensation would occur
	for $d>3$, while the discussion is more involved in $d=3$ because the
	critical temperature vanishes.

	\section{The partition function}
	\label{sec:partition-function}
	
	The partition function of a massless scalar field on the spatial manifold
	$\mathbb R^{d-1}\times \mathbb S_{L}$ with $d>1$
	\begin{equation}
		\label{eq:7}
		Z(\beta,\mu)={\rm Tr}\,e^{-\beta(\hat H-\mu\hat D)},
	\end{equation}
	may be computed by operator methods. In terms of the dimensionless
	parameters $b=\frac{\beta}{L}$, $u=\mu\frac{L}{2\pi}$ and the volume
	$V_{d-1}$ of $\mathbb R^{d-1}$, the exact result takes the
	form (details of the computations can be found in \cite{ Aggarwal:2024axv, Alessio:2021krn})
	\begin{equation}
		\label{eq:8}
		\ln Z(\beta,\mu)=\frac{V_{d-1}}{L^{d-1}}\big[\xi(d+1)b+\xi(d)\frac{1}{b^{d-1}}
		+l(\beta,\mu)\big],
	\end{equation}
	where $\xi(z)=\pi^{-\frac z2}\Gamma(\frac z2)\zeta(z)$. The first term
	is directly related to the Casimir energy of the system. The second
	term comes from the modes with $n=0$ and coincides with the
	contribution of a massless scalar on the spatial manifold
	$\mathbb R^{d-1}$. In the Casimir case, it comes from photons that
	propagate parallel to the plates. The part of interest to us here is
	the last one,
	\begin{equation}
		\label{eq:9}
		l(\beta,\mu)=-\frac{2\pi^{\frac{d-1}{2}}}{\Gamma(\frac{d-1}{2})}
		\sum_{n\in\mathbb{N}^*}\sum_{\pm}\int^\infty_0 d\kappa \kappa^{d-2} \ln{(1-e^{-2\pi b[\sqrt{\kappa^2+ n^2}\mp u]})},
	\end{equation}
	which is singular for $|u|> 1$. Therefore, the maximum value that
	$|\mu|$ can take is $ 2\pi \over L $. It follows that the chemical
	potential vanishes in the large volume limit, $L\rightarrow \infty $,
	and the effect we are studying is a finite sized effect. The function
	$l(\beta,\mu) $ may be expressed in terms of polylogarithms \cite{MR618278, srivastava2011zeta} as
	\begin{equation}
		\label{eq:15}
		l(\beta,\mu)=\frac{1}{(4\pi)^{\frac{d-1}{2}}b^{d-1}}\sum_{n\in \mathbb
			N^*}\sum_{\pm}\sum_{k=0}^\infty
		\frac{\Gamma(\frac{d+1}{2}+k)}{\Gamma(\frac{d+1}{2}-k)k!}(4\pi
		bn)^{\frac{d-1}{2}-k}\, {\rm Li}_{\frac{d+1}{2}+k}(e^{-(1\mp\frac{u}{n})2\pi bn}),
	\end{equation}
	where the sum over $k$ cuts at $\frac{d+1}{2}-1$ for odd $d$. The
	derived quantities of interest are the ``charge density''
	$\delta(\beta,\mu)$ defined by
	\begin{equation}
		\langle\hat D \rangle=\frac 1\beta\partial_\mu\ln
		Z(\beta,\mu)=\frac{V_{d-1}}{L^{d-1}}\delta(\beta,\mu)\label{eq:24},
	\end{equation}
	and the Casimir pressure 
	\begin{equation}
		\label{eq:22}
		p=\frac{1}{V_{d-1}}\partial_{L}\big[\beta^{-1}\ln Z(\beta,\mu)\big]. 
	\end{equation}
	In the case of photons, this last result has to be multiplied by $2$
	since $L=2a$. Equation \eqref{eq:8} is directly adapted to a
	low temperature/small box expansion $b\gg 1$, where below criticality
	$|u|<1$, the leading exponentially suppressed contributions are given
	by
	\begin{equation}
		\label{eq:16}
		l(\beta,\mu)=\sum_\pm \frac{1}{b^{\frac{d-1}{2}}}{\rm
			Li}_{\frac{d+1}{2}}
		(e^{-(1\mp u)2\pi b})+\dots.
	\end{equation}
	
	Using standard results, the partition function may also be written in
	a form adapted to a high temperature/large box expansion $b\ll 1$ as
	\begin{multline}
		\label{eq:13}
		\ln Z(\beta,\mu)=\frac{V_{d-1}}{L^{d-1}}
		\Big[\frac{\Gamma(\frac{d+1}{2})}{2\pi^{\frac{d+1}{2}}b^{d}}\sum_\pm
		{\rm Li}_{d+1}(e^{\pm 2\pi
			bu})+\frac{\Gamma(\frac{d}{2})}{2\pi^{\frac{d}{2}}b^{d-1}}
		\big[2\zeta(d)
		-\sum_\pm
		{\rm Li}_{d}(e^{\pm 2\pi bu})\big]\\+m(\mu)+h(\beta,\mu)\Big],
	\end{multline}
	where
	\begin{equation}
		\label{eq:27}
		m(\mu)=2\sum_{p\in \mathbb
			N^*}\big(\frac{iu}{p}\big)^{\frac{d}{2}}K_{\frac d2}(2\pi
		iup)=(iu)^{\frac{d-1}{2}}\sum_{k=0}^\infty
		\frac{\Gamma(\frac{d+1}{2}+k){\rm
				Li}_{k+\frac{d+1}{2}}(e^{-2\pi i
				u})}{\Gamma(\frac{d+1}{2}-k)k!(4\pi i u)^{k}},
	\end{equation}
	with $m(0)=\xi(d)$ and where the sum over $k$ cuts at
	$k=\frac{d-1}{2}$ for odd $d$ \cite{Haber:1981tr}.
	This result may also be expressed in terms of generalized Clausen
	functions using
	\begin{equation}
		{\rm Li}_{s}(e^{-2\pi i u})={\rm C}_{s}(2\pi u)-i{\rm
			S}_{s}(2\pi u)\label{eq:28}. 
	\end{equation}
	The exponentially suppressed terms are 
	\begin{equation}
		\label{eq:23}
		h(\beta,\mu)=\frac{2}{b^{\frac d2}}\sum_{\pm,p,n \in \mathbb
			N^*}\big(\frac{n\pm iub}{p}
		\big)^{\frac d2} K_{\frac d2}(\frac{2\pi(n\pm iub)p}{b}).
	\end{equation}
	The expansion of the terms in the first  line of \eqref{eq:13} may
	be obtained from
	\begin{equation}
		\label{eq:10}
		{\rm Li}_n(e^{\nu})=\frac{\nu^{n-1}}{(n-1)!}\big[H_{n-1}-\ln (-\nu)\big]+
		\sum_{k=0,k\neq n-1}^\infty\frac{\zeta(n-k)}{k!}\nu^k,
	\end{equation}
	for $n\in \mathbb N^*$ and $|\nu|<2\pi$. Here the harmonic number is
	$H_n=\sum_{h=1}^n\frac{1}{h}$ with $H_0=0$.

	\section{Bose-Einstein condensation of scalar field model in higher
		dimensions}
	\label{sec:bose-einst-cond}
	
	Bose Einstein condensation of the massless scalar field model occurs in spatial dimensions
	$d>3$ at the critical values $|u|=1$.
	
	In the low temperature/small box regime, it follows from \eqref{eq:16}
	that the critical temperature is
	\begin{equation}
		\label{eq:20}
		T_C=\frac{1}{L}\Big(\frac{|\delta|}{\zeta(\frac{d-1}{2})}\Big)^{\frac{2}{d-1}}, 
	\end{equation}
	where the sign of $\delta$ follows that of the critical value. The
	charge density $\delta_G$ of the ground state is
	\begin{equation}
		\label{eq:21}
		\left\{  \begin{array}{l}T>T_C:  \delta_G=0,\\
			T<T_C:
			\frac{\delta_G}{\delta}=1-(\frac{T}{T_C})^{\frac{d-1}{2}}
			\quad{\rm and}\ u=\pm 1. 
		\end{array}\right.
	\end{equation}
	
	In a high temperature/large box regime, when taking into account
	\eqref{eq:13} and \eqref{eq:10}, the leading contribution to the
	charge density is
	\begin{equation}
		\label{eq:31}
		\delta(\beta,\mu)
		\approx \frac{(d-1)\xi(d-1)u}{b^{d-1}},
	\end{equation}
	while critical temperature and charge density of the
	ground state now become 
	\begin{equation}
		\label{eq:33}
		T_C=\frac{1}{L}\Big(\frac{|\delta|}{(d-1)\xi(d-1)}\Big)^{\frac{1}{d-1}},
	\end{equation}
	respectively
	\begin{equation}
		\label{eq:21a}
		\left\{  \begin{array}{l}T>T_C:  \delta_G=0,\\
			T<T_C: \frac{\delta_G}{\delta}=1-(\frac{T}{T_C})^{d-1}
			\quad{\rm and}\ u=\pm 1. 
		\end{array}\right.
	\end{equation}
	
	\section{Critical behavior in three dimensions}
	\label{sec:behav-at-crit}
	
	In the physical dimension $d=3$, which is the case relevant for
	photons in a Casimir box, the charge density diverges logarithmically
	at criticality $|u|=1$. This implies in particular that the critical
	temperature $T_C$ for Bose-Einstein condensation vanishes.
	
	(i) In the high temperature/large box regime $b\ll 1$, the 
	expansion of \eqref{eq:13} yields the
	corrections to the black body result,
	\begin{multline}
		\label{eq:30}
		\ln Z(\beta,\mu)=\frac{V_2}{L^2}\Big[
		\frac{\zeta(4)}{\pi^2b^3}+\frac{2 u^2\zeta(2)}{b}+\pi u^2 \ln b\\+\pi
		u^2\big[-\frac{3}{2}+\ln (2\pi |u|)\big]+\frac{C_3(2\pi u)}{2\pi}+uS_2(2\pi
		u)+\mathcal{O}(b)
		\Big]+\dots.
	\end{multline}
	
	The signature of the phase transition appears in the b-independent term of the expansion given by the second line of \eqref{eq:30}. Indeed, the charge
	density becomes
	\begin{equation}
		\label{eq:17}
		\delta(\beta,\mu)=\frac{2u\zeta(2)}{\pi b^2}+\frac{u\ln b}{b}+
		\frac{\iota(\mu)}{b}+\mathcal{O}(b)+\dots ,
	\end{equation}
	where
	\begin{equation}
		\label{eq:32}
		\iota(\mu)=u\Big[\ln \big|\frac{\pi u}{\sin
			(\pi u)}\Big|-1\Big],
	\end{equation}
	diverges logarithmically at criticality $|u|=1$.
	
	\begin{figure}[h!]
		\centering
		\includegraphics[scale=0.80]{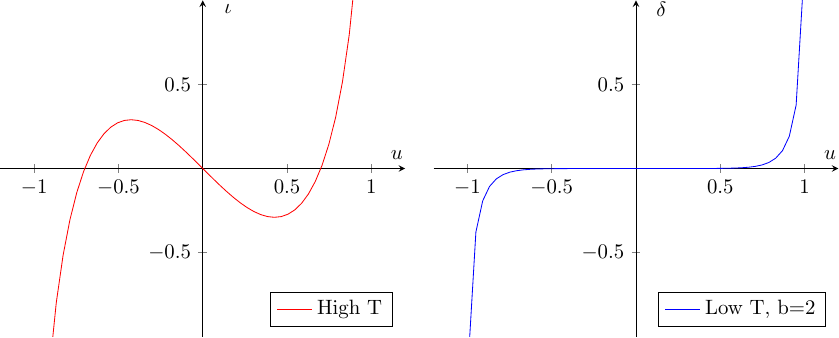}
		\caption{\label{fig:iota} Logarithmic divergence at high and low temperature}
	\end{figure}
	
	(ii) In the low temperature/small box regime $b\gg 1$, it
	follows from \eqref{eq:15} that
	\begin{equation}
		\label{eq:11}
		l(\beta,\mu)=\frac{1}{4\pi b^2}\sum_\pm \Big[(4\pi b){\rm
			Li}_2(e^{-(1\mp u)2\pi b})
		+2{\rm Li}_3(e^{-(1\mp u)2\pi b})\Big]+\dots, 
	\end{equation}
	where the dots denote exponentially suppressed contributions. The
	charge density,
	\begin{equation}
		\label{eq:19}
		\delta(\beta,\mu)=\frac{1}{b}\ln\Big[\frac{1-e^{-(1+u)2\pi
				b}}{1-e^{-(1-u)2\pi b}}\Big]+\mathcal{O}(b^{-2})\dots,
	\end{equation}
	again diverges logarithmically at criticality. 
	
	Furthermore, this is accompanied by a quantum phase
	transition. Indeed, $l(\beta,\mu)$ is itself exponentially suppressed when
	$|u|<1$. At criticality $|u|=1$ however, this exponential suppression
	turns into a power law. More precisely, the non critical phase with
	exponential suppression corresponds to $|u|$ sufficiently far from $1$
	so that $(1-|u|)2\pi b\gg 1$, while the two critical phases correspond
	to $|u|$ sufficiently close to $1$ so that $(1-|u|)2\pi b\ll 1$. The
	cross-over regions are around the lines $(1-|u|)2\pi b=1$. In the
	critical phases, one may use \eqref{eq:10} for the expansion of the
	polylogarithms,
	\begin{multline}
		\label{eq:18}
		l(\beta,\mu)=\frac{1}{b}\Big[\zeta(2)|u|+(|u|-1)2\pi b\\-(|u|-1)2\pi
		b\ln((1-|u|)2\pi b)+\mathcal{O}((|u|-1)2\pi
		b)^2\Big]+\mathcal{O}(b^{-2})+\dots. 
	\end{multline}
	\\
	\begin{figure}
		\centering
		\includegraphics{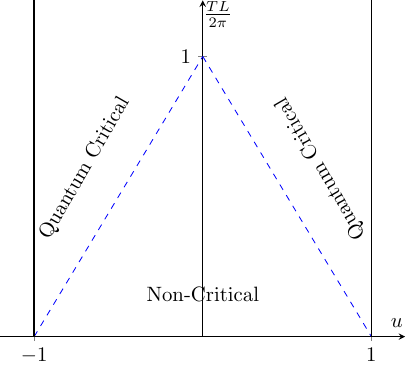}
		\caption{\label{fig:qc} Cross-over regions of the quantum
			phase transition}
	\end{figure}
	The interpretation is as follows. At low temperature/small box, the
	system is a non-interacting collection in $2$ large spatial
	dimensions of a massless scalar and complex massive scalars of
	increasing mass $m_{n}=\frac{2\pi}{L}n$. The singular contribution to
	the partition function is dominated by the complex scalar with the
	lowest mass at $n=1$, which determines the critical behavior in this
	regime.

	\section{Gravitons}
	\label{sec:gravitons}
	
	The considerations above generalize directly to the case of gravitons
	with suitably defined perfectly conducting boundary conditions because
	the partition function can be shown to be the identical to that of photons
	\cite{Alessio2021}. This equivalence means every result derived in earlier sections of the paper, i.e., the closed-form partition function, the systematic high and low temperature expansions, and, most importantly, the complete critical analysis in the physical case d=3, transfers immediately and rigorously to gravitons. Below, we will just provide the boundary conditions and the mapping to the electromagnetic problem worked out in all details in~\cite{Alessio2021}.  In the simple context of free fields, this
	phase transition for gravitons is a concrete realization of some of
	the ideas put forward in \cite{Dvali:2011aa,Dvali:2012en}. 
	
	The geometric setup is same as in section~\ref{sec:spectrum}. The difference is
	that here we work with linearised gravity (around flat space) with the
	following boundary conditions for gravitational perturbations, $ h_{ij} $,
	and their conjugate momenta, $ \pi^{ij} $,
	\begin{equation}\label{key}
		h_{ab}|_{\partial V}=0=\pi^{ab}|_{\partial V}, \quad h_{33}|_{\partial V}=0=\pi^{33}|_{\partial V}, \quad \partial_3 h_{a3}|_{\partial V}=0=\partial_3 \pi^{a3}|_{\partial V}~.
	\end{equation}
	The linear polarization tensors for polarizations `$+ $' and `$\times$' are
	\begin{align}
		e_{TT+}^{ab}&=\frac{1}{\sqrt{2}k_{\perp}^2k^2}\left (k^2\epsilon^{ac}k_c\epsilon^{bd}k_d-k_3^2k^ak^b\right ), & e_{TT\times}^{ab}&=\frac{k_3}{\sqrt{2}k_{\perp}^2k}\left (\epsilon^{ac}k_ck^b+k^a\epsilon^{bc}k_c\right ), \cr
		e_{TT+}^{a3}&=\frac{1}{\sqrt{2}k^2}k^ak^3,&e_{TT\times }^{a3}&=-\frac{1}{\sqrt{2}k} \epsilon^{ac}k_c, \cr
		e_{TT+}^{33}&=-\frac{1}{\sqrt{2}k^2}k_{\perp}^2, &e_{TT\times}^{33}&=0~.
	\end{align}
	Here $TT$ refers to the transverse-traceless components of the fields,
	$ h_{ij}, \pi^{ij} $ that turn out to be the physical modes.
	Defining the oscillator variables
	\begin{equation}
		\label{eq:1}
		a_k^{TT s}=\frac 12\sqrt k h^{TT s}_k+i \frac{\pi^{TT s}_k}{\sqrt k},\quad s=(\times,+),
	\end{equation}
	the problem can be mapped on to electromagnetism in a Casimir box
	upon the identification
	\begin{equation}
		a_k^{TT\times}\rightarrow	a^E_k, \quad a_k^{TT+}\rightarrow a^H_k~.
	\end{equation}
	
	\paragraph{Note:} As in the electromagnetic case, the boundary conditions \eqref{key} are to be
	understood as idealized, perfectly-reflecting conditions on the metric
	perturbations. We use them here as a toy model to study the influence of
	boundary conditions in linearized gravity before turning to more physically
	relevant boundary conditions like those discussed in \cite{garfinkle1985possibility} or in
	\cite{1978MNRAS.185..833Z,Damour:1978cg} (extended from electromagnetism to
	linearized gravity).
	
	\section{Discussion}
	\label{sec:discussion}
	
	From a theoretical viewpoint, the exact results derived here are closely related
	to those that appear when generalizing modular invariance from two
	\cite{Polchinski:1985zf,Itzykson1986} to higher spacetime dimensions
	\cite{Cappelli:1988vw,Cardy:1991kr,Shaghoulian:2015kta,Berg:2019jhh,Alessio:2021krn,Aggarwal:2024axv},
	combined with the technique of going back to real rather than purely imaginary
	chemical potential \cite{ROBERGE1986734,Alford:1998sd,PhysRevD.75.025003}.
	
	The next step consists in studying the effects of adding background curvature
	and interactions. This should give rise to a more interesting Bose-Einstein
	condensate and will allow one to study spin effects, that is to say to
	distinguish between massless scalar fields, photons and gravitons. How the
	effects discussed here are modified when moving away from idealized boundary
	conditions to realistic metal plates, as reviewed for instance in chapters 13
	and 14 of \cite{Bordag:2009zzd}, should also be most instructive.
	
	\section*{Acknowledgments}
	\label{sec:acknowledgements}
	
	\addcontentsline{toc}{section}{Acknowledgments}
	
	The authors are grateful to F.~Alessio and M.~Bonte for discussions and
	collaboration on closely connected subjects. This work is supported by the Fund
	for Scientific Research (F.R.S.-FNRS) Belgium through a research fellowship for
	AA and conventions FRFC PDR T.1025.14 and IISN 4.4503.15. AA was partially supported by the Austrian Science Fund (FWF), projects P~32581, P~33789, and P~36619.
	
	\providecommand{\href}[2]{#2}\begingroup\raggedright\endgroup

	%\bibliography{master.bib}
\end{document}